\documentclass[english,aps,tightenlines,showpacs, showkeys, notitlepage]{revtex4-2}
\usepackage[T1]{fontenc}
\usepackage[latin9]{inputenc}
\usepackage{color}
\usepackage{babel}
\usepackage{amstext}
\usepackage{graphicx}
\usepackage{esint}
\usepackage[unicode=true,pdfusetitle,
 bookmarks=true,bookmarksnumbered=false,bookmarksopen=false,
 breaklinks=false,pdfborder={0 0 0},pdfborderstyle={},backref=false,colorlinks=true]
 {hyperref}
\begin{document}
\title{Cosmological implications of Born--Infeld-$f(R)$ gravity}
\author{Salih Kibaro\u{g}lu$^{1,2}$}
\email{salihkibaroglu@maltepe.edu.tr}

\author{Emilio Elizalde$^{2}$}
\email{elizalde@ice.csic.es}

\date{\today}
\begin{abstract}
A modified Born-Infeld gravitation theory with a $f\left(R\right)$
function being added to the determinant action is analyzed from a
cosmological viewpoint. The corresponding accelerating dynamics are
studied in a simplified conformal approach without matter. Three different
structures for the auxiliary metric function are analyzed, with the
aim to establish a deeper understanding of the role of this function
in cosmology. After performing the analysis, it is seen that, by modifying
the auxiliary metric function, a Big Rip singularity or either a Little
Rip dark energy model may arise. 
\end{abstract}
\affiliation{$^{1}$Maltepe University, Faculty of Engineering and Natural Sciences,
34857, Istanbul, Turkey}
\affiliation{$^{2}$Institute of Space Sciences (CSIC-IEEC) C. Can Magrans s/n,
08193 Cerdanyola (Barcelona) Spain}
\maketitle

\section{Introduction}

General relativity (GR) has been proven to be an extremely successful
theory. It is now, in a word, the standard theory to describes the
dynamics of the gravitational field at all scales. However, despite
its impressive success, this theory has been seen to be unable to
explain gravitational phenomena at very extreme conditions, in particular
at very high energies, where quantum effects do play a critical role.
This is invoked as one of the main motivations for studying alternative
theories of gravity, in the search for better fitting theories valid
in these extreme conditions. One should here recall, on passing, that
Einstein himself (as well as a number of noted theoreticians, subsequently)
was already convinced, when he constructed it, that his theory would
need to be improved, namely, that it was not in its final form. In
view of that, the idea we are pursuing here is not that extraordinary
or new.

Good examples of departures from GR are the different Born--Infeld
inspired modifications of gravity. In analogy with the Born-Infeld
theory for nonlinear electromagnetism \citep{born1934foundations},
in 1998 Deser and Gibbons established a gravitational model by using
Born-Infeld like determinant structures involving the Ricci tensor,
instead of the electromagnetic field tensor \citep{deser1998born}.
This study remained, however, unsuccessful because of the appearance
in the process of ghost-like instabilities. To overcome this unwanted
situation, Vollick analyzed the idea from a different point of view,
and succeeded to construct a theory without ghosts. He did this by
using the Palatini formulation, in which the connection and the metric
tensor are taken to be independent fields \citep{vollick2004palatini,vollick2005born}.
In 2010, Bañados and Ferreira improved such theory, endowing it with
a more general structure, in particular concerning the definition
of the matter term \citep{Banados2010eddington}. 

As a follow up to these works, this theory has continued to attract
a great deal of attention, including numerous applications in cosmology,
astrophysics, and many other uses in the literature (for a detailed
review, see \citep{jimenez2018born}). 

In the Palatini formulation, it is hard to find a consistent generalization
of the Born-Infeld gravity, because this one is known to have some
restrictions and constraints to the theory. In 2014, a non-perturbative
and consistent generalization was obtained by combining Born-Infeld
gravity with an $f\left(R\right)$ function in the framework of the
Palatini formalism \citep{makarenko2014born}, namely

\begin{equation}
S_{BI}=\frac{1}{\kappa^{2}\epsilon}\int d^{4}x\left[\sqrt{-|g_{\mu\nu}+\epsilon R_{\mu\nu}\left(\Gamma\right)|}-\lambda\sqrt{-|g_{\mu\nu}|}\right]+\frac{\alpha}{2\kappa^{2}}\int d^{4}x\sqrt{-g}f\left(R\right)+S_{m},\label{eq: BI-fR}
\end{equation}
where $g_{\mu\nu}$ is the space-time metric, while $R_{\mu\nu}\left(\Gamma\right)$
corresponds to the Ricci tensor of the connection, which is here fully
independent of the metric, $\lambda$ is a constant, $f\left(R\right)$
a function of the Ricci scalar $R=g^{\mu\nu}R_{\mu\nu}\left(\Gamma\right)$,
and $S_{m}$ is the matter action, which depends on the metric field,
only.

This model above gained a lot of attention and it is still used in
a variety of ranges of fields in cosmology, in particular within dark
energy models \citep{odintsov2014born,Makarenko:2014nca,makarenko2014unification,Makarenko:2014cca,Elizalde:2016vsd,chen2016modified,banik2018dynamical}.
As is well known, Palatini $f\left(R\right)$ theories can be solved
by introducing an auxiliary metric, which is conformal with the space-time
metric $g_{\mu\nu}$ (for details see, e.g., the review \citep{Olmo:2011uz}).

In the present paper, we analyze the auxiliary metric function by
applying the techniques given in \citep{Makarenko:2014nca}, and we
assume that it has an exponential structure with respect to time.
Unlike the well-known version given in Eq.(\ref{eq: BI-fR}), we are
here mainly interested in the model where the $f\left(R\right)$ term
directly enters into the determinant structure \citep{makarenko2014born}.
This model has been much less studied than the other one. It will
be seen here that it is capable to provide a quite different, and
on its turn, very interesting framework \citep{chen2016modified}.
For this purpose, we start by examining the evolution of the universe
in the context of dark energy scenarios for several different forms
of the auxiliary metric.

The paper is organized as follows. In Section II, we provide a brief
review of Born-Infeld-$f\left(R\right)$ gravity. Section III is devoted
to the study of the effects of the auxiliary metric on the model,
leading to different possibilities. Finally, the last section contains
the conclusions of the paper and a final discussion.

\section{Born-Infeld-$f\left(R\right)$ gravity}

Let us briefly review the Born-Infeld-$f\left(R\right)$ theory with
a function of the Ricci scalar being added to the determinant action
\citep{makarenko2014born} (see also \citep{Makarenko:2014nca,odintsov2014born,makarenko2014unification}).
To avoid any ghost instabilities, the theory is formulated in the
framework of the Palatini formalism (for more detail see \citep{Olmo:2011uz,olmo2012open}),
in which the metric $g_{\mu\nu}$ and the connection $\Gamma_{\beta\gamma}^{\alpha}$
are treated as independent variables. This model has been partly analyzed
in \citep{makarenko2014born,chen2016modified} (for early development
similar to this model, in the pure metric formalism, see \citep{comelli2005determinant},
and in the teleparallel framework, see \citep{Fiorini:2013kba}).
We use the construction method for the Born-Infeld inspired action
given in \citep{chen2016modified} to definitely realize our purpose.
With this purpose, we start with the following action, 

\begin{eqnarray}
S_{BI} & = & \frac{2}{\kappa}\int d^{4}x\left[\sqrt{-|g_{\mu\nu}+\kappa F_{\mu\nu}|}-\lambda\sqrt{-|g_{\mu\nu}|}\right]+S_{m},\label{eq: action_f(r)_2}
\end{eqnarray}
where $\kappa$ is a dimensional constant and the symmetric tensor
$F_{\mu\nu}$ is defined as,

\begin{equation}
F_{\mu\nu}=\alpha R_{\mu\nu}\left(\Gamma\right)+\beta g_{\mu\nu}F\left(R\right),
\end{equation}
where $F\left(R\right)$ is a function of the Ricci scalar, $\alpha$
and $\beta$ are dimensionless constants. 

To find the field equations, we first define a new symmetric object
as
\begin{equation}
p_{\mu\nu}=g_{\mu\nu}+\kappa\left[\alpha R_{\mu\nu}+\beta g_{\mu\nu}F\left(R\right)\right],\label{eq: p_f(R)}
\end{equation}
where the inverse of $p_{\mu\nu}$ is denoted $\left(p^{-1}\right)^{\mu\nu}$
and these objects satisfy $\left(p^{-1}\right)^{\mu\rho}p_{\rho\nu}=\gamma_{\nu}^{\mu}$.
By using Eq.(\ref{eq: p_f(R)}), the action Eq.(\ref{eq: action_f(r)_2})
becomes
\begin{eqnarray}
S_{BI} & = & \frac{2}{\kappa}\int d^{4}x\left[\sqrt{-p}-\lambda\sqrt{-g}+\frac{\kappa}{2}\mathcal{L}_{m}\right],
\end{eqnarray}
where $p=det\left(p_{\mu\nu}\right)$. To obtain equations of motion,
we look for the variation with respect to the metric $g_{\mu\nu}$
and the connection $\Gamma_{\beta\mu}^{\alpha}$, respectively. Those
are

\begin{eqnarray}
\sqrt{-p}\left(p^{-1}\right)^{\mu\nu}\left(1+\kappa\beta F\left(R\right)\right)-\kappa\beta\sqrt{-p}\left(p^{-1}\right)^{\sigma\rho}g_{\sigma\rho}F_{R}g^{\mu\alpha}g^{\nu\beta}R_{\alpha\beta}-\lambda\sqrt{-g}g^{\mu\nu} & = & -\kappa\sqrt{-g}T^{\mu\nu},\label{eq: eom_g-2}
\end{eqnarray}
and

\begin{equation}
\nabla_{\lambda}\left[\sqrt{-p}\left(\alpha\left(p^{-1}\right)^{\mu\nu}+\beta\left(p^{-1}\right)^{\sigma\rho}g_{\sigma\rho}g^{\mu\nu}F_{R}\right)\right]=0.\label{eq: eom_conn-1}
\end{equation}
As we mention before, we know that the Palatini $f\left(R\right)$
gravitation theories can be solved by using an auxiliary metric which
is conformal with the space-time metric. But in the Born-Infeld-$f\left(R\right)$
action this assumption is actually not enough and the connection equation
Eq.(\ref{eq: eom_conn-1}) cannot actually be solved by using the
$p_{\mu\nu}$ tensor \citep{makarenko2014born}. Working on this idea
and taking moreover into account Eq.(\ref{eq: eom_conn-1}), we then
define an auxiliary metric $u_{\mu\nu}$ as follows 

\begin{equation}
\sqrt{-p}\left(\alpha\left(p^{-1}\right)^{\mu\nu}+\beta\left(p^{-1}\right)^{\sigma\rho}g_{\sigma\rho}g^{\mu\nu}f_{R}\right)=\sqrt{-u}\left(u^{-1}\right)^{\mu\nu},\label{eq: def_u}
\end{equation}
where $u=det\left(u_{\mu\nu}\right)$ and $\left(u^{-1}\right)^{\mu\nu}$
is the inverse of $u_{\mu\nu}$, that is $\left(u^{-1}\right)^{\mu\rho}u_{\rho\nu}=\delta_{\nu}^{\mu}$.
By considering the determinant of Eq.(\ref{eq: def_u}), we find,
\begin{equation}
p^{2}det\left(\alpha\left(p^{-1}\right)^{\mu\nu}+\beta\left(p^{-1}\right)^{\sigma\rho}g_{\sigma\rho}g^{\mu\nu}F_{R}\right)=u,
\end{equation}
and this definition leads us to obtain
\begin{eqnarray}
\left(u^{-1}\right)^{\mu\nu} & = & \frac{\left(\alpha p^{\mu\nu}+\beta p^{\sigma\rho}g_{\sigma\rho}g^{\mu\nu}F_{R}\right)}{\sqrt{det\left(\alpha\delta_{\xi}^{\zeta}+\beta p^{\sigma\rho}g_{\sigma\rho}p_{\xi}^{\zeta}F_{R}\right)}}.\label{eq: u_munu}
\end{eqnarray}
Now Eq.(\ref{eq: eom_conn-1}) can be rewritten as
\begin{equation}
\nabla_{\lambda}\left[\sqrt{-u}\left(u^{-1}\right)^{\mu\nu}\right]=0,\label{eq: eom_u-1}
\end{equation}
which is in analogy with Einstein's theory, in which the connection
equation takes the form $\nabla_{\lambda}\left[\sqrt{-g}g^{\mu\nu}\right]=0$,
in the torsion-less case (for more details see \citep{Misner:1973prb}).
Therefore, we can describe the connection in terms of the auxiliary
metric $u_{\mu\nu}$, in the following form: 

\begin{equation}
\Gamma_{\mu\nu}^{\rho}=\frac{1}{2}\left(u^{-1}\right)^{\rho\sigma}\left(u_{\sigma\nu,\mu}+u_{\mu\sigma,\nu}-u_{\mu\nu,\sigma}\right).\label{eq: gamma_u}
\end{equation}

\subsection{The conformal case}

We can provide the main definition for this case if we assume the
following conformal relationship between $p_{\mu\nu}$ and $g_{\mu\nu}$
\begin{equation}
p_{\mu\nu}=f\left(t\right)g_{\mu\nu},\label{eq: p_conf}
\end{equation}
then Eq.(\ref{eq: def_u}) takes the form,
\begin{equation}
f\left(t\right)\left(\alpha+4\beta F_{R}\right)\sqrt{-g}g^{\mu\nu}=\sqrt{-u}\left(u^{-1}\right)^{\mu\nu}.
\end{equation}
From this definition, we can easily write
\begin{equation}
u_{\mu\nu}=f\left(t\right)\left(\alpha+4\beta F_{R}\right)g_{\mu\nu}.\label{eq: func_u}
\end{equation}
In this case, the conditions are satisfied, as given in Eq.(\ref{eq: eom_u-1})
and Eq.(\ref{eq: gamma_u}). The conformal case (\ref{eq: p_conf})
leads to write a corresponding condition, in which the Ricci tensor
is proportional to the metric tensor $g_{\mu\nu}$ as,
\begin{equation}
R_{\mu\nu}=r\left(t\right)g_{\mu\nu},\label{eq: Ricci_g_relation}
\end{equation}
where $r\left(t\right)$ can be found by using Eq.(\ref{eq: p_f(R)})
and Eq.(\ref{eq: p_conf}), to be
\begin{equation}
r\left(t\right)=\frac{1}{\kappa\alpha}\left[f\left(t\right)-1-\kappa\beta F\left(R\right)\right].\label{eq: r_2}
\end{equation}
Furthermore, taking trace of Eq.(\ref{eq: Ricci_g_relation}), one
can also find that $r\left(t\right)=\frac{R}{4}$ and then Eq.(\ref{eq: Ricci_g_relation})
reduces to
\begin{equation}
R_{\mu\nu}=\frac{R}{4}g_{\mu\nu}.
\end{equation}
To develop a cosmological scenario, let us consider a homogeneous
and isotropic case, namely the Friedman--Lemaitre--Robertson--Walker
(FLRW) universe, with the well-known metric 
\begin{equation}
ds^{2}=-dt^{2}+a^{2}\left(t\right)\left(dx^{2}+dy^{2}+dz^{2}\right),\label{eq: metric_FRW}
\end{equation}
where $t$ is the cosmic time and $a\left(t\right)$ is the scale
factor. Now, we can define the auxiliary metric, as
\begin{equation}
u_{\mu\nu}=u\left(t\right)\text{diag}\left(-1,a^{2}\left(t\right),a^{2}\left(t\right),a^{2}\left(t\right)\right),\label{eq: metric_U}
\end{equation}
where $u\left(t\right)=f\left(t\right)\left(\alpha+4\beta F_{R}\right)$.
From this background, one can find the following expressions
\begin{equation}
r\left(t\right)=3\left(H+\frac{\dot{u}}{2u}\right)^{2},\label{eq: r_1}
\end{equation}
\begin{equation}
2\dot{H}=H\frac{\dot{u}}{u}+\frac{3}{2}\left(\frac{\dot{u}}{u}\right)^{2}-\frac{\ddot{u}}{u},
\end{equation}
where the upper dot denotes the time derivative and $H=\frac{\dot{a}}{a}$
is the Hubble parameter. 

Using these equations, one can derive the following relation
\begin{equation}
u\left(t\right)=cr\left(t\right),\label{eq: u_r_relation}
\end{equation}
where $c$ is an integration constant. Then, combining Eq.(\ref{eq: r_1})
and Eq.(\ref{eq: u_r_relation}), one gets
\begin{equation}
H=\pm\sqrt{\frac{u}{3c}}-\frac{\dot{u}}{2u}.\label{eq: Hubble_parameter}
\end{equation}
Furthermore, with the help of Eqs.(\ref{eq: u_r_relation}) and (\ref{eq: func_u}),
we find the $F\left(R\right)$ function is given in the following
form

\begin{equation}
F\left(R\right)=-\frac{1}{\kappa\beta}-\frac{\alpha}{4\beta}R\pm\frac{\sqrt{cR^{2}+\frac{1}{c_{1}^{2}\kappa^{2}\beta^{2}}}}{4\kappa\beta},\label{eq: F(R)-1}
\end{equation}
where $c_{1}$ is an integration constant. 

Now, substituting Eq.(\ref{eq: F(R)-1}) into (\ref{eq: action_f(r)_2})
and settings $c=16$ and $\lambda=\frac{1}{16c_{1}^{2}\kappa^{2}\beta^{2}}$,
the action takes finally the the form, 

\begin{equation}
S=\frac{2}{\kappa}\int d^{4}x\sqrt{-g}R^{2}.\label{eq: action_2}
\end{equation}

\section{Analysis of the auxiliary metric function}

\begin{figure}
\includegraphics[width=8cm]{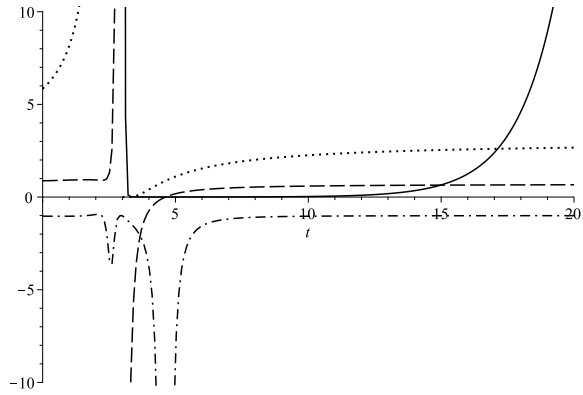}

\caption{Considering positive sign in (\ref{eq: Hubble_parameter}) and the
case (\ref{eq: f(t)_1}), this figure shows $u\left(t\right)$ (dotted
line), $a\left(t\right)$ (solid line) in Eq.(\ref{eq: a(t)}), $H\left(t\right)$
(dashed line) in Eq.(\ref{eq: H(t)}) and $w_{eff}\left(t\right)$
(dash-dot line) in Eq.(\ref{eq: w(t)}) with the parameters $a_{0}=0.001$,
$u_{0}=3$, $c=2$, $n=-1$, $h=2$ and $t_{0}=3$.\label{fig: big_rip_1}}
\end{figure}

In the paper \citep{Makarenko:2014nca}, the authors show that any
kind of dark energy cosmology may be derived from Born-Infeld-$f\left(R\right)$
theory by assuming different structures for the auxiliary metric function.
From this point of view, in this section, we will investigate the
exponential characteristics of the metric function $u\left(t\right)$
to find its effects on the acceleration dynamics of the ensuing cosmology.
For a general situation, when we solve Eq.(\ref{eq: Hubble_parameter})
with respect to $a\left(t\right)$, we get the evolution of the scale
factor as
\begin{equation}
a=a_{0}e^{\int\left(\pm\sqrt{\frac{u}{3c}}-\frac{\dot{u}}{2u}\right)dt}.\label{eq: a_general}
\end{equation}
Then using Eq.(\ref{eq: Hubble_parameter}), one can find the general
expression of the effective equation of state (EoS) as (for more detail
see \citep{nojiri2007introduction,nojiri2011unified,Nojiri:2017Modified}),
\begin{equation}
w_{eff}=-1-\frac{2\dot{H}}{3H^{2}}=-1+\frac{4\left(\pm\sqrt{\frac{3u}{c}}u\dot{u}+3\dot{u}^{2}-3u\ddot{u}\right)}{3\left(-\frac{4u^{3}}{c}\pm3\sqrt{\frac{3u}{c}}u\dot{u}-3\dot{u}^{2}\right)}.
\end{equation}
Let us now have a closer look at our main problem here. We assume
the $u\left(t\right)$ function to have a exponential characteristic
with respect to time, as follows 

\begin{equation}
u=u_{0}e^{f\left(t\right)},\label{eq: u(t)}
\end{equation}
where $f\left(t\right)$ is an arbitrary function of time and $u_{0}$
is an arbitrary constant. From this definition, the scale parameter
can be found by inserting Eq.(\ref{eq: u(t)}) into Eq.(\ref{eq: a_general}),
as follows,
\begin{equation}
a=a_{0}e^{\int\left(\pm\sqrt{\frac{u_{0}}{3c}}e^{\frac{f}{2}}-\frac{\dot{f}}{2}\right)dt},\label{eq: a(t)}
\end{equation}
where $a_{0}$ is an integration constant and the Hubble function
in Eq.(\ref{eq: Hubble_parameter}) reads

\begin{equation}
H=\pm\sqrt{\frac{u_{0}}{3c}}e^{\frac{f}{2}}-\frac{\dot{f}}{2}.\label{eq: H(t)}
\end{equation}
Finally, the effective EoS parameter becomes

\begin{equation}
w_{eff}=-1\mp\frac{\left(4\sqrt{3u_{0}c}\dot{f}e^{\frac{f\left(t\right)}{2}}-6c\ddot{f}\right)}{\left(3\sqrt{c}\dot{f}\pm2\sqrt{3u_{0}}e^{\frac{f}{2}}\right)^{2}}.\label{eq: w(t)}
\end{equation}
In this regard, the condition $f\left(t\right)=ht$ has been analyzed
in \citep{Makarenko:2014nca} where $h$ is a constant. Within certain
condition, this assumption leads to a Little Rip universe \citep{frampton2011little,frampton2012models,brevik2011viscous,brevik2012turbulence,makarenko2013big}
($a\left(t\right)\rightarrow\infty$ and $H\left(t\right)\rightarrow\infty$
at future infinity). In our first case, we take this assumption a
step further, to the following form
\begin{equation}
f\left(t\right)=h\left(t_{0}-t\right)^{n},\label{eq: f(t)_1}
\end{equation}
where $t_{0}$ and $n$ are constants. The time-dependence of the
scale factor, Hubble parameter, effective EoS parameter and the auxiliary
metric function, respectively, is drawn in Fig.(\ref{fig: big_rip_1}).
According to this graph, the Big Rip singularity \citep{Caldwell:2003vq,Nojiri:2003vn,Faraoni:2001tq}
occurs at the instant $t=t_{0}$. Moreover, we observe that the scale
factor has a bouncing characteristic and we also note that, in this
case, the effective EoS parameter at this time is equal to minus one.

\begin{figure}
\includegraphics[width=8cm]{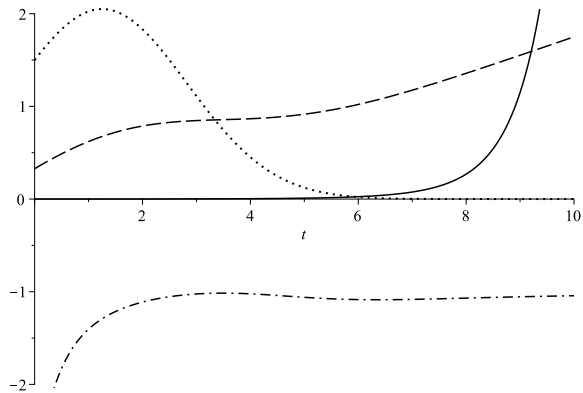}

\caption{Considering positive sign in (\ref{eq: Hubble_parameter}) and the
case (\ref{eq: k(t)_1}), this figure shows $u\left(t\right)$ (dotted
line), $a\left(t\right)$ (solid line) in Eq.(\ref{eq: scale_factor_1}),
$H\left(t\right)$ (dashed line) in Eq.(\ref{eq: hubble_1}) and $w_{eff}\left(t\right)$
(dash-dot line) in Eq.(\ref{eq: EoS_1}) with the parameters $a_{0}=0.0005$,
$u_{0}=1.5$, $c=1.5$, $n=2$, $h=0.5$ and $q=-0.2$.\label{fig: little_rip_1} }
\end{figure}

Now we extend this assumption with the following expression
\begin{equation}
f\left(t\right)=ht+k\left(t\right),\label{eq: f(t)_2}
\end{equation}
where $k\left(t\right)$ is an time-dependent function. For this background,
we derive the scale factor, the Hubble parameter and the effective
EoS, respectively, as follows

\begin{equation}
a=a_{0}e^{\int\left[\pm\sqrt{\frac{u_{0}}{3c}}e^{\frac{ht+k}{2}}-\frac{1}{2}\left(h+\dot{k}\right)\right]dt},\label{eq: scale_factor_1}
\end{equation}

\begin{equation}
H=\pm\sqrt{\frac{u_{0}}{3c}}e^{\frac{ht+k}{2}}-\frac{1}{2}\left(h+\dot{k}\right),\label{eq: hubble_1}
\end{equation}

\begin{equation}
w_{eff}=-1\mp\frac{\left[4\sqrt{3u_{0}c}\left(h+\dot{k}\right)e^{\frac{ht+k}{2}}-6c\ddot{k}\right]}{\left[3\sqrt{c}\left(h+\dot{k}\right)\pm2\sqrt{3u_{0}}e^{\frac{ht+k}{2}}\right]^{2}}.\label{eq: EoS_1}
\end{equation}
From this assumption, the second case is characterized by the following
form
\begin{equation}
k\left(t\right)=qt^{n},\label{eq: k(t)_1}
\end{equation}
where $q$ is a constant and time-dependence of $a\left(t\right)$,
$H\left(t\right)$, $w_{eff}\left(t\right)$ and $u\left(t\right)$
are presented in Fig.(\ref{fig: little_rip_1}). At this point, we
see that the scale factor and the Hubble parameter go to infinite
at future infinity and the effective EoS less than $-1$ and it asymptotically
approaches $-1$. According to these results, one can say that this
model obeys the rules of the Little Rip dark energy scenarios. 

\begin{figure}
\includegraphics[width=8cm]{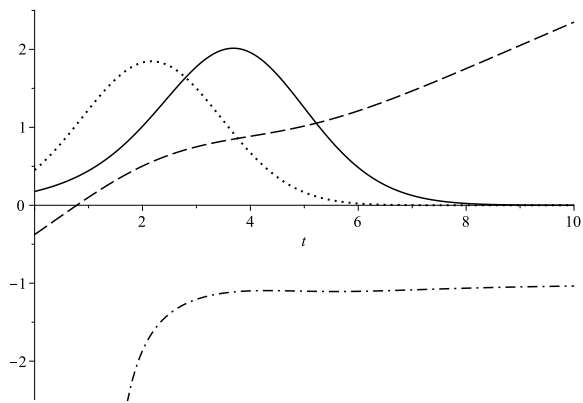}

\caption{Considering positive sign in (\ref{eq: Hubble_parameter}) and the
case (\ref{eq: k(t)_2}), this figure shows $u\left(t\right)$ (dotted
line), $a\left(t\right)$ (solid line) in Eq.(\ref{eq: scale_factor_1}),
$H\left(t\right)$ (dashed line) in Eq.(\ref{eq: hubble_1}) and $w_{eff}\left(t\right)$
(dash-dot line) in Eq.(\ref{eq: EoS_1}) with the parameters $a_{0}=0.7$,
$u_{0}=1.5$, $c=2$, $n=2$, $t_{0}=2$, $h=0.1$ and $q=-0.3$.\label{fig: Third_cond}}
\end{figure}

Finally, as our third case, we consider the following function:
\begin{equation}
k\left(t\right)=q\left(t_{0}-t\right)^{n}.\label{eq: k(t)_2}
\end{equation}
We plot the functions $a\left(t\right)$, $H\left(t\right)$, $w_{eff}\left(t\right)$
and $u\left(t\right)$ for this case in Fig.(\ref{fig: Third_cond}).
In this case, we see that the scale factor starts with an expansion
and after reaching a maximum value it continues with a contraction
process. Besides, the Hubble parameter goes to infinite for infinite
time and the effective EoS does not cross the phantom barrier in future
times and asymptotically reaches the value $-1$, similar as in the
case of the Little Rip universe.

\section{Conclusion}

In this work, we have considered the cosmological implications of
Born-Infeld-$f\left(R\right)$ gravity without matter where the $f\left(R\right)$
term enters into the square root in the Palatini formulation. To this
aim, we have thrown a closer look into the auxiliary metric function
$u\left(t\right)$. In this regard, assuming that it has an exponential
structure with respect to time, Eq.(\ref{eq: u(t)}), and we have
examined the auxiliary metric function in three different cases.

In the first case, Eq.(\ref{eq: f(t)_1}), we have obtained a Big
Rip singularity at $t=t_{0}$. In addition, we observed that the scale
factor has a bouncing characteristic, which starts as a contraction
and continues as an expansion. Regarding the effective equation of
state, when $t=t_{0}$ its value is $-1$ and after a period of time
it asymptotically closes to $-1$. In this case one can affirm that
it is possible to get other types of finite-time future singularities
\citep{nojiri2005properties}. 

In the second case, Eq.(\ref{eq: k(t)_1}), we have reached the Little
Rip universe, which displays non-singular characteristics. Finally,
in the last case, Eq.(\ref{eq: k(t)_2}), we have obtained a kind
of closed universe in which the scale factor starts with an expansion
and, after reaching a maximum value, it begins to contract. In the
contraction period, it asymptotically closes to zero. Besides, the
Hubble parameter goes to infinite at future infinity, the effective
EoS behaves like in the Little Rip model, and the metric function
reaches its maximum value when $t=t_{0}$.

According to these results, we have shown that the Born-Infeld-$f\left(R\right)$
theory within the conformal approximation provides indeed a useful
background to study dark energy scenarios, simply by modifying the
auxiliary metric function. The nature of dark energy, which is thought
to be responsible for the late time accelerated expansion of the universe
\citep{perlmutter1998discovery,perlmutter1999measurements,Spergel:2003elm,Spergel:2006bqn},
still remains a mystery. In this respect, and according to our findings
here, we may conclude that the class of Born-Infeld-$f\left(R\right)$
theories have good potential to help understanding this very challenging
problem.
\begin{acknowledgments}
The authors wish to thank Sergei D. Odintsov for useful discussions
and comments regarding the results presented in this work. This work
has been partially supported by MICINN (Spain), project PID2019-104397GB-I00,
of the Spanish State Research Agency program AEI/10.13039/501100011033,
by the Catalan Government, AGAUR project 2017-SGR-247, and by the
program Unidad de Excelencia María de Maeztu CEX2020-001058-M. S.K.
is supported by the Scientific and Technological Research Council
of Turkey (TUB\.{I}TAK) under grant number 2219-A.
\end{acknowledgments}

\bibliographystyle{apsrev4-2}
\bibliography{BIFR2_dark}

\end{document}